\RequirePackage[2021-01-01]{latexrelease}
\documentclass{llncs}

\usepackage{vdm-rg}
\usepackage{graphicx}
\def\ProofIndent{0em}
\renewcommand{\VDMInnerIndent}{\quad}
\newcommand{\I}[1]{\hbox to #1em{}}
\title{Using Rely/Guarantee to Pinpoint Assumptions underlying Security Protocols}
\author{Nisansala P. Yatapanage\inst{1}\and%\orcidID{0000-0002-0498-513} \and
Cliff B. Jones\inst{2}%\orcidID{0000-0002-0038-6623}}
}
\institute{School of Computing, The Australian National University, Australia \and
School of Computing, 
Newcastle University,
NE4 5TG, UK
}

\authorrunning{Nisansala P. Yatapanage and Cliff B. Jones}

\begin{document}

\maketitle

\noindent 
\makebox[\linewidth]{\today} \\
% comment out to restore standard runningheads
%\newcommand{\version}{draft --- ready to show to friends}
%\newcommand{\kopf}{\textnormal{\today.\ Version\ \version}}
\pagestyle{myheadings}
%\markboth{\kopf}{\kopf}
%
%\fbox{Version: \version}\hfill
%\fbox{Dated: \today}


\begin{abstract}
This paper explores the application of a state-based specification notation (VDM) and associated rely-guarantee ideas to security.
%added a sentence to link to Overture
The verification of security protocols is essential in order to ensure the
absence of potential attacks. However, verification results are only valid with
respect to the assumptions under which the verification was performed. 
These assumptions are often difficult to identify and hidden in documentation, making it unclear
whether a given protocol is safe to deploy in a particular environment.
Rely-guarantee ideas offer a way to reason abstractly about 
interference from an environment:
using this approach, assumptions can be recorded and
made precise. This paper investigates this approach on the
Needham-Schroeder Public Key protocol and Lowe's extension, showing that the technique can
effectively uncover 
--and provide a way to record-- 
the assumptions under which the protocol can withstand attacks from intruders.
\end{abstract}

%\discuss{can we explore ways of avoiding ``anthropomorphic'' terms?\\
%something like ``has evidence for'' might work better than ``believe''? \\
%(NPY) Sure, if you can think of how to say that. I started to replace ``believed partner"
%below, but you still have to say ``the user whom they have evidence \ldots" so there's still a \textit{they} in there. I wouldn't have a clue how to get rid of it without the sentence becoming
%really long and hard to read. Personally, I don't think we should go to great lengths to avoid anthropomorphic terms at the expense of it becoming hard to read, and it's not a far stretch to consider
%the users as people, since they are \textit{users}, who could be people. If you have a nice way of changing it, though, I'm happy with that.\\
%(C) ``the user for which there is evidence \ldots" \\
%how about $intPartner$ (for ``intended'')?
%}
%
%\issue{The ToC is temporary but it might help tidy up titles}
%{for example:\\
%\S\ref{S-props} probably ought move to either \S\ref{S-intro} or \ref{S-overall-spec}\\
%and \S\ref{S-postpone-encryption}'s title/position is somehow wrong?}

%\setcounter{tocdepth}{2}
%\tableofcontents
%\newpage

%%%%%%%%%%%%%%%%%%%%%%%%%
\section{Introduction}
\label{S-intro}

This paper is an experiment in specification and reasoning styles for security protocols;
these are certainly not solved problems.
Most protocol descriptions actually start with a sketch of the intended communications between ``principals'';
attacks are commonly handled as {\em post facto} test cases that show how specific properties might be violated ---
but these desirable properties are rarely made clear as a pre-specification.
Moreover,
protocols
--even corrected ones--
do not come with recorded assumptions about the behaviour of unintended
(or undesired)
participants.
For example,
in Lowe's attack~\cite{lowe1995attack} on the Needham-Schroeder~\cite{NeedhamSchroeder-78}
protocol,
the nonces appear to be a means-to-an-end 
(of authentication)
but are then sometimes used as session keys.
Even in~\cite{BoydEt-protocols-19}
the roles of Authentication and Key Establishment (AKE) are not always distinct.

Protocol specification cannot be achieved simply:
standard pre/post specifications record things that should be achieved by the conclusion of executions,
but leakage of information is about situations that should not occur during execution.
Rely-guarantee conditions 
(see below)
were proposed to facilitate compositional development of shared-variable concurrent software
but have subsequently been shown 
to be useful for recording assumptions about external interference from components that are not under the control of designers.

Assumptions are fundamental to establishing the viability of security protocols.
This invites the question of whether rely-guarantee conditions can be useful in the formal specification and justification of security protocols;
specifically, for recording the assumptions made about attackers and 
proving protocol safety with respect to the assumptions.
This paper explores evidence for a positive answer to the question. 
Furthermore, it is argued that challenging the assumptions provides a method for incorporating run-time checks 
that facilitate a reduction of assumptions about the behaviour of miscreants in protocols.

%\sep

It is well known that the original Needham-Schroeder (N-S) protocol~\cite{NeedhamSchroeder-78} is ``flawed'';
but we want to identify 
(and record) 
a set of assumptions which might have resulted in the residual flaw;
furthermore, we consider Gavin Lowe's ``corrected'' protocol~\cite{lowe1995attack} 
and identify assumptions that appear to be made in justifying his suggested correction.

Perhaps more importantly,
what protocol designers (including Needham, Schroeder and Lowe) try to do is to build checks into their designs
to reduce the assumptions that are made about the behaviours of intended and unintended participants.
Here again,
clear recording of assumptions is imperative.
Examples of assumptions include:

\begin{enumerate}

\item secret keys are never shared

\item brute force attacks are not economic

\item if $\alpha$ sends a secret under the public key of $\beta$,
$\beta$ will not leak the original secret.

\item if $\alpha$ sends a secret under the public key of $\beta$,
receipt of that same secret
(under the public key of $\alpha$) 
justifies the conclusion that any accompanying new secret must have been generated by $\beta$.

\item a process is communicating with its intended recipient.

\end{enumerate}

\noindent
%\plannote{(NPY)We had the following sentence here: \\
% I see your point - I have reworded
The first two of these relate narrowly to encryption;
the other assumptions also relate to the content of the messages sent. 

%I took them out because I think even the other assumptions relate to encryption and the content.
%If you want to keep this sentence, maybe reword it?}

Being explicit about the assumptions is desirable because it invites discussion 
(and pre-deployment confirmation);
it can also expose attack vectors;
furthermore, it can help designers to increase the checking in a protocol
and such checks can be shown to yield protocols that need less dangerous assumptions.
Above all,
if the claim is that a protocol manages to reduce the assumptions by adding checks,
it must be worth making the assumptions explicit;
furthermore,
there should be a record of the argument as to correctness under the stated assumptions.

%\plan{To get down to conference page limits, these 2 subsections will have to be a sentence each (pointing to sources)}
%\plan{Also need to figure out just how much we need to say here
%origin~\cite{Jones81d}
%recent (algebraic view)~\cite{HJM-23}
%application to CPS~\cite{BurnsHayesJones-19} 
%}

%%%%%%%%%%
\subsection{Rely-guarantee thinking}
\label{S-RG}

%\plan{Probably we will use nested R/G to discuss when the protocol aborts so that needs a brief description as well
%\cite{JonesBurns-23b}}
Space limitations make it impossible to describe all of the background material.
%\plan{Brief outline of VDM notation used (but we'll try to minimise so write later)}
%Similarly,
%for page-limit reasons, 
The long-established VDM notation is not described here but can be studied in~\cite{Jones90a};
there is also an international standard on VDM~\cite{VDMSL}.

A convenient publication outlining the rely-guarantee ideas is~\cite{HayesJones18};
early papers on rely/guarantee ideas include~\cite{Jones81d,Jones83a,Jones83b};
an algebraic presentation and more recent references are covered in~\cite{HJM-23}.
%\plannote{(C) add on separate systems\\
%(and if used) layered R/G}
R-G was conceived as a way of decomposing a specified component into sub-components that executed concurrently.
It was subsequently observed 
(e.g.~\cite{JonesHJ07})
that rely conditions could be used to describe assumptions about external components that were not being designed:
they were part of the problem space.
In a further development
(e.g.~\cite{BurnsJones:ECRTS:2022})
it was shown how layered R-G conditions could be used to describe fault-tolerant systems:
optimistic rely conditions are linked to optimal behaviour but useful behaviour can still be achieved under weaker assumptions.

\section{Protocols}
\label{S-props}

In order to verify any protocol, it is first necessary to determine what properties it is required to fulfil
and, crucially, what assumptions have been made.
As the specification is developed, the assumptions required for the protocol to satisfy these potential properties
should be identified.

\subsection{The Needham-Schroeder protocol proposal}
\label{S-N-S}

%The general aim of protocols \ldots
Most protocol discussions start by listing the (expected) interactions;
for example the public key version of Needham-Schroeder (N-S) is normally presented as~\cite{NeedhamSchroeder-78}:

\begin{formula}
(a1) A: enc([A, NA], pkeym(B))\\
(b1) B: enc([NA, NB], pkeym(A))\\
(a2) A: enc([NB],  pkeym(B))
\end{formula}

\noindent
User $A$ initiates the protocol by sending a message ($a1$) to some other participant, $B$. 
This message,
encrypted with the public key of $B$,
contains $A$'s identity and a fresh nonce, $NA$.
The intended recipient, $B$, replies with a message $b1$ 
encrypted with the public key of the user whose identity was in message $a1$ (i.e.~$A$)
containing both $NA$, as confirmation of having
read the $a1$ message and a new nonce, $NB$.
Finally, $A$ replies with $a2$
(encrypted with the public key of $B$)
containing $NB$ to confirm receipt of $b1$.
In addition,
there are implied aborts in situations where the expected nonces in messages $b1$ or $a2$ are absent.

A train of thought about this exchange might be that only $B$ can decrypt $a1$ and therefore $A$
has evidence%
\footnote{We avoid terms such as ``$A$ believes \ldots'' because such protocols are algorithms being followed as defined;
one might choose to say that ``a designer believes \ldots''.}
in message $b1$ both that $B$ received $a1$ and that $NB$ was created by $B$;
similarly $a2$ provides evidence to $B$ that $A$ received $b1$.

Clearly,
encryption provides some level of security against attack
but it is important to identify how little can be concluded:
for example, $B$ has no evidence from message $a1$ as to who actually sent it or the uniqueness of $NA$;
whilst $A$ has evidence from receipt of $b1$ that $B$ 
(or someone with $B$'s secret key)
must have decrypted message $a1$,
this does not establish that no one else now has $NA$.
Apart from the lack of specification/assumptions,
it is not made explicit under what circumstances the protocol should abort.

The objective of the current paper is to make the specifications of such protocols precise ---
in particular, recording assumptions that are required in order to verify that the protocols satisfy their specifications.

\subsection{Protocol specification}
\label{S-protocols}

A significant and recurrent problem is the lack of precise specifications for protocols in the literature. 
In both the original paper by Needham and Schroeder and Lowe's version \cite{lowe1995attack},
the overall goal of the protocol is never actually stated, 
thus making it difficult to employ formal verification.

At first it might appear that the aim is to allow two participants to share secrets that are invisible to all other users. 
However, this cannot be the case, as there is no way to ensure the secrecy of the messages unless both participants are guaranteed to conform to the protocol. 
In the presence of dishonest users, a user who legitimately receives a message can forward it to anyone else.%
\footnote{This seems to contrast with Lowe's claim that the shared message nonces can be used for authentication purposes for further transactions; perhaps Lowe's intent was that the shared nonces be used in conjunction with the confirmed identities of the participants.}
This could be stated as:

\begin{conjecture} \label{prop-secrets} % We know this is false, but we should get to that later.
If a protocol session between users $X$ and $Y$ has terminated successfully, and the nonces used for the
protocol were $NX$ and $NY$, then only $X$ and $Y$ know both $NX$ and $NY$.
\end{conjecture}

As an alternative,
it might be assumed that the aim of the N-S protocol is instead to provide some form of authentication, i.e. allowing each participant to confirm the identity of the other participant with whom they are communicating. 
This would appear to be the view of~\cite[\S4.3.3]{BoydEt-protocols-19} and could be stated as:

%A key property of the N-S protocol 
%%(see \S\ref{S-N-S})
%appears to be authentication:
%upon successful termination of the protocol,
%the identities of both participants have been confirmed:

\begin{conjecture} \label{prop-auth}
If a protocol session between users $X$ and $Y$ has terminated successfully, then $Y$ has confirmed the identity
of $X$, and vice versa.
\end{conjecture}

\noindent
Note however that at this stage it is not yet clear what ``has confirmed the identity'' means. 
This needs to be determined 
as the specification is developed.

As will be seen in the following sections, these conjectures only hold under strong assumptions. In the case of Conjecture \ref{prop-secrets}, the 
required assumptions are unrealistic and this property therefore does not hold in practice for the un-modified N-S protocol
(nor, in fact, the Lowe correction).

\subsection*{Abstract models of protocols}

Any abstraction leaves out certain aspects (e.g. hiding ``side-channel'' attacks), while exposing
other aspects. The question is whether a given model exposes the required aspects. 
The current authors wonder whether using process algebra channels can inadvertently hide assumptions.
The temptation to base specification and verification on process algebra
(CSP is used by Lowe in~\cite{lowe1996breaking})
might be dangerous because hiding information on channels is precisely the problem that has to be solved.

%\plannote{(C) I've read more on Lowe:
%do we want to say more?}

%\plan{I think (with some checking) we could expand the paragraph above to include at least the ``spi-calculus''?
%My view is that any process algebra has baked in point-to-point communication which is precisely what the protocols are trying to establish}

The goal of the rely/guarantee approach is precisely to identify and record assumptions.
In our model,
all messages are visible to every user,%
\footnote{This makes it easy to model ``man in the middle'' attacks;
see, for example, the discussion in the next section.}
but, if properly encrypted,
only the intended recipient can decrypt a message.
This follows the principles of the Dolev-Yao intruder model \cite{DolevYao83}, often used in security protocol verification.

%%%%%%%%%%%%%%%%%%%%
\section{Optimistic specification (postponing encryption)}
\label{S-postpone-encryption}

Faced with the question,
``what is the spec?'',
we start with a rather optimistic one: 
after some set of actions, two users each record that the other is their partner
and they share two nonces which no one else knows (Conjectures \ref{prop-secrets} and \ref{prop-auth}). Under \emph{strong} 
(i.e. optimistic and uncheckable) 
assumptions, the original N-S protocol achieves this. Tackling this case exemplifies our approach 
and makes a systematic questioning of the assumptions possible.
%
%\issue{Do we say here that encryption gets treated in \S\ref{S-reif}?}

The intention is to postpone discussing encryption and to treat it as an abstraction:
a message,
$mk-Msg(u\sb{r}, u\sb{s}, vl)$, where $u\sb{r}$ is the receiver and $u\sb{s}$ is the sender,
 is assumed to be only readable by $u\sb{r}$.
Employing this abstraction exposes important assumptions about 
the flow of information between users in the N-S protocol.%
\footnote{There is no claim that this is an original idea --- it is for example used in~\cite{sprenger2010developing}.}
Encryption as a way to achieve this abstraction is covered in \S\ref{S-reif}.
%\footnote{Compare with our paper on ``separation as an
% abstraction''~\cite{JonesYatapanage-15}.}
%Removed the next line as it was repetitive (we said three times that it's postponed to S5!).
%and the conditions under which the assumptions can be achieved by encryption are postponed to Section~\ref{S-reif}.
Postponing encryption has an additional notational advantage that the reasoning in this section is not complicated by terms of the form $dec(enc(\cdots))$.

%%%%%%%%%%
\subsection{Abstract state}
\label{S-abs-spec}

Messages are modelled as a record containing the $Uid$ of the intended recipient, the sender's $Uid$ and the content of the message.
\begin{record}{Msg}
      rec: Uid\\
      sender: Uid \\
      content: \seqof{Item}
\end{record}

All users can access all $Msg$s,
so they are all in the $history$ component of the state ($\Sigma$);
the $rec$ field governs who can access the $content$.
%\plan{The $sender$ field is a ghost variable for use in proofs.}
The $sender$ field is a ``ghost variable'' that is used in correctness arguments;
it is not to be confused with data included in the $content$ field that can be a $Uid$
(but might be fraudulently set).

\type{Item}{Nonce | Uid}

Considering only one protocol exchange can be confusing since it is possible for any user process to be involved in multiple overlapping ``sessions''.
One example is where a ``man in the middle'' uses an innocent receiver to generate nonces 
that are replayed to an initiating user of another session.
%in this way the intruder can obtain full knowledge of the nonces that appear to be confidential to the user processes that do follow the protocol.
For this reason, 
the models below separate information by session identifiers and record the origin of nonces.

%\plannote{(NPY) Cliff, I took out your sentence about the intruder getting full knowledge because that isn't correct: the
%man in the middle attack doesn't allow the intruder to gain knowledge apart from what was given to them anyway - it's only
%Lowe's attack that gives the attacker extra information.}

%\plannote{By all means reword
%but, without some statement of why the effect of the two sessions is bad news for the good guys, it's not clear what is amiss \\
%(NPY) But it isn't bad news for the good guys if there's a man-in-the-middle attack - it's only
%Lowe's attack that achieves a bad outcome. Do you want to mention something like that instead?}

%\plannote{(C) I noticed that we don't use the $complete$ field in a meaningful way in this section;
%unless this changes, I suggest that we remove it?}

Information about $User$s is stored in:%
\footnote{The secret key fields ($skey: SKey$) and public keys ($PKey$) are discussed in Section~\ref{S-reif}.}

\begin{record}{User}
	intPartner: \mapof{Sid}{Uid}\\
	knows: \mapof{Sid}{\setof{Nonce}} \\
	skey: SKey\\
    	conforms: \Bool\\
    	complete: \mapof{Sid}{\Bool}
\end{record}

%\plannote{(C) I believe it is a theorem (of the $conforms$ code) that the first element of any $knows$ pair = $intPartner$\\
%do you agree?
%(NPY) Yes, sorry, after thinking about it more, I realised I was wrong yesterday and you're right - we can remove the first
%element in the pairs.}

\noindent
User records keep track of the intended partner and whether or not the session was successful ($complete$ is $\false$ at the beginning of a session).
If user $x$ is playing the role of $A$ in the N-S protocol, then $\sigma.users(x).intPartner$
 contains the $Uid$ of the user with whom $x$ intended to communicate
 (i.e.~the user for whom $A$ encrypted $a1$). 
 If the user $y$ is playing the
 role of $B$ in the protocol, then $\sigma.users(y).intPartner$ contains the $Uid$ of the user with whom $y$ will communicate
 (i.e.~the $Uid$ given as part of the $a1$ message received by $B$). The $knows$ field keeps track of the 
 set of nonces that a user knows, either by inventing the nonce or receiving it in a message.

Assumptions about the behaviour of other users could be recorded as rely conditions but 
it is more convenient to make them invariants.
Note that 
--as in~\cite{JonesBurns-23b}--
preservation of the invariant by any context becomes an extra assumption as though it were part of every rely condition 
and an extra obligation as though it were conjoined to every guarantee condition.

%\plannote{(C) It would be easier for the reader if we stuck to ``definition before use''
%if you agree, there are some more things that need moving?}

The notion of $Action$ includes both messages and invention of nonces:

\type{Action}{Msg | Invent}

\begin{record}{Invent}
      user: Uid  \hspace{10em} \\
      what: Nonce
\end{record} 

The following function yields the history restricted to a single user's activities.

\begin{fn}{uHist}{as,user} \\ 
\signature{\seqof{Action} \x Uid \to \seqof{Action}}
\If as = \emptyseq
\Then \emptyseq
\Else{
 \If \hd{as} \in Msg \And ((\hd{as}).rec = user \Or (\hd{as}).sender = user) 
 \Then \seq{\hd{as}} \sconc uHist(\tl{as}, user)
 \Else{
   \If \hd{as} \in Invent \And (\hd{as}).user = user 
   \Then \seq{\hd{as}} \sconc uHist(\tl{as}, user)
   \Else uHist(\tl{as}, user)
   \Fi}
 \Fi}
\Fi
\end{fn}

To record the overall intent of the protocol,
the most abstract specification is based on a state:

\begin{record}{\Sigma}
  users: \mapof{Uid}{User} \\
  history: \seqof{Action}\\
  pkeys: \mapof{Uid}{PKey}% \hspace{4em} \hbox{but postpone encryption}
 \end{record}
 \where
\begin{fn}{inv-\Sigma}{\sigma}\\
   unique-nonces(\sigma.history) \And 
   no-read-others(\sigma) \And \\
   \forall*{u \in \dom{\sigma.users}}{
   \sigma.users(u).conforms \Implies \R
   no-leaks(uHist(\sigma.history,u)) \And no-forge(uHist(\sigma.history,u))}
\end{fn}

The predicates used in $inv-\Sigma$ are now defined.
$Nonce$s are assumed to be ``fresh'':%
\footnote{This assumption is not treated formally below but actually depends on the low probability of a clash in a sufficiently large space.}

\begin{fn}{unique-nonces}{as}\\
\signature{\seqof{Action} \to \Bool}
\forall*{i, j \in \inds{as}}{
\set{as(i), as(j)} \subseteq Invent \And
 as(i).what = as(j).what \Implies i = j}
\end{fn}

The predicate $no-read-others$ records the assumption that no user can read the contents of a message other than the intended
recipient (given by the $rec$ field of the message). 
This is an abstract version of the idea of encryption, which is explored in Section~\ref{S-reif}.

 \begin{fn}{no-read-others}{\sigma}
 \signature{\Sigma \to \Bool} \\
 \forall*{u \in \dom \sigma.users}{
   \forall*{n \in \sigma.users(u).knows}{\exists*{m \in \rng \sigma.history}{
     	m \in Invent \, \And \, m.user = u \, \And \, m.what = n  \Or \\
   	m \in Msg \, \And \, m.rec = u \, \And \, n \in \elems{m.content} }}
   }
 \end{fn}

A predicate that limits sending of non-locally generated nonces must, however,  allow their return to the believed originator.
There are two ways of tackling this:
the simpler employs the ghost variable field $sender$, which has to be a ghost variable since no user process can actually
determine which user sent a $Msg$.

\begin{fn}{no-leaks}{as}\\
\signature{\seqof{Action}  \to \Bool}
\forall*{i, j \in \inds{as}}{
    i < j \Implies \\
    	\Let mk-Msg(\beta, \alpha, vl\sb{i}) = as(i) \In
    	\Let mk-Msg(\gamma, \beta, vl\sb{j}) = as(j) \In
      (\exists{c \in Nonce} {c \in \elems{vl\sb{i} \inter \elems{vl\sb{j}}}) \Implies \gamma = \alpha}}
\end{fn}

\noindent
It is however the case that no user process can ensure that $no-leaks$ is satisfied precisely because it requires knowledge of the ghost field $sender$.
An alternative $no-app-leaks$ could use the claimed sender field instead:

\begin{fn}{no-app-leaks}{as}\\
\signature{\seqof{Action}  \to \Bool}
\forall*{i, j \in \inds{as}}{
    i < j \Implies \\
    	\Let mk-Msg(\beta, , vl\sb{i}) = as(i) \In
    	\Let mk-Msg(\gamma, , vl\sb{j}) = as(j) \In
      \exists*{\alpha \in Uid, c \in Nonce}{
      	\alpha \in \elems{vl\sb{i}} \And c \in \elems{vl\sb{i} \inter \elems{vl\sb{j}}} \Implies 
	\gamma = \alpha}}
\end{fn}

\noindent
but this requires all messages to be signed.
Whilst this is true in Lowe's extension
(see Section~\ref{S-Lowe-attack}),
it is not the case in the original N-S protocol.%
%\footnote{It is worth noting that Lowe's attack
%(see Section~\ref{S-Lowe-attack})
%still works under the assumption of $no-app-leaks$.}

A predicate for ensuring that users only sign honestly is:

\begin{fn}{no-forge}{as} \\ 
\signature{\seqof{Action} \to \Bool}
 \forall*{mk-Msg(\beta, \alpha, vl) \in \elems{as}}{
		\forall*{i \in \inds{vl}}{ vl(i) \in Uid \Implies	vl(i) = \alpha}}
\end{fn}

%\issue{NB: $history$ is not used in $NS$,
%I think we can remove it and the $rely/guar$}

The following dynamic invariant also holds on the states:
this requires that nothing can change an $Action$ that has occurred,
a user's conformity cannot change
and that users retain all the information that was added to $knows$.

 \begin{fn}{dyn-inv}{\sigma, \sigma'}\\
 \signature{\Sigma \x \Sigma \to \Bool}
         \sigma. history \subseteq \sigma'.history \And\\
	 \forall* {u \in \sigma.users}{\forall* {sess \in \dom \sigma.knows(sess)} {
	 	\sigma.users(u).knows(sess) \subseteq \sigma'.users(u).knows(sess) \And \\
	 	\sigma'.users(u).conforms = \sigma.users(u).conforms}}
 \end{fn}

%
%\begin{figure}
%\begin{op}[NS]
%\args{from: Uid, to: Uid, sess: Sid}
%\ext{\Wr users: \mapof{Uid}{User}        }
%\pre{\Not users(from).complete(sess) \And \Not users(to).complete(sess)}
%\rely{ \set{from, to} \dres users' = \set{from, to} \dres users}
%\guar{ \set{from, to} \dsub users' = \set{from, to} \dsub users}
%\post{users'(from).complete(sess) \And users'(to).complete(sess) \And \\
%users'(from).intPartner(sess) = to \And \\
%		  users'(to).intPartner(sess) = from \And \\
%         \exists*{NA, NB \in Nonce}{
%         \set{NA, NB} \subseteq users'(from).knows(sess) \And\\
%         \set{NA, NB} \subseteq users'(to).knows(sess) \And\\
%          \forall{u \in (\dom{users} \minus \set{from, to})}{
%        	  	 users'(u).knows(sess) \, \inter \, \set{NA, NB} = \emptyset
%        	}}
%      }
%\end{op} 
%\caption{Specification for Needham-Schroeder (relying on optimistic assumptions)}
%\label{F-NS-spec}
%\end{figure}

%%%%%%%%%%
\subsection{Specification}

An optimistic overall objective can be captured by the specification given in Fig.~\ref{F-spec-ideal}.
The argument is that N-S and the NS(L) code satisfies this strong specification and that Lowe's extension
(see Section \ref{S-Lowe}) also does something sensible (aborts) in the case of Lowe's attack.

\begin{figure}
%\plannote{N/C agreed to make both $s1/s2$ arguments\\
%that would make it worth changing $noMods$ to be for one $Uid$}

%\plannote{(C) Oh dear! at some point we have lost the key property that no other user has $NA/NB$!
%pls check my addition \\
%(NPY) I agree in theory, but the problem is that do we say $noleaks$ at this level? It doesn't seem to be
%in the invariant anymore! Without it, your addition doesn't hold.}

\begin{op}[NS]
\args{from: Uid, to: Uid, sf: Sid, st: Sid}
\ext{\Wr users: \mapof{Uid}{User} \\
        \Wr history: \seqof{Action}}
\pre{
\Not users(from).complete(sf) \And \Not users(to).complete(st) \And\\
users(from).conforms \And users(to).conforms}
\rely{noMods(users, users', from, sf) \And noMods(users, users', to, st)}
\guar{noModsToOthers(users, users', \set{from, to}, sf) \, \And \\ noModsToOthers(users, users', \set{from, to}, st)}
\post{users'(from).intPartner(sf) = to \And\\
         users'(to).intPartner(st) = from\And \\
         users'(from).complete(sf) \And         	
         users'(to).complete(st) \And\\
         \exists*{NA,NB \in Nonce}{
                  NA \neq NB \, \And \\
                  \set{NA, NB} \subseteq users'(from).knows(sf)  \And\\
                  \set{NA, NB} \subseteq users'(to).knows(st) \And\\
                  \Not \exists*{u \in (Uid \minus \set{from, to}), su \in Sid}{
                  	\set{NA, NB} \subseteq users'(u).knows(su)}}
      }
\end{op} 

%Fig \ref{F-spec-ideal} uses the following functions:

\begin{fn}{noMods}{userm, userm', u, sess} \\ 
\signature{(\mapof{Uid}{User}) \x (\mapof{Uid}{User}) \x Uid \x Sid \to \Bool}
userm'(u).complete(sess) = userm(u).complete(sess) \And \\
userm'(u).intPartner(sess) = userm(u).intPartner(sess) 
\end{fn}

%\plannote{Why can't the next be simplified? \\
%(NPY) Because I was trying to say that if a user is not in $leave$ then nothing is modified,
%but if they are in $leave$, we still promise not to modify the user's other sessions. I think
%your new version says we can modify a user's other sessions if they are in $leave$. \\
%Also, the word ``leave" is confusing - isn't it the opposite? The ones in $leave$ are the ones we
%can modify, not the ones we're leaving alone.}
%
%\begin{fn}{noModsToOthers}{userm, userm', good} \\ 
%\signature{(\mapof{Uid}{User}) \x (\mapof{Uid}{User}) \x \setof{Uid} \to \Bool}
%(\forall {u \in (\dom{userm} \minus good)}{ userm'(u) = userm(u)} ) %\And \\
%%\set{leave} \dsub userm'(u).complete = \set{leave} \dsub userm(u).complete \And \\
%%\set{leave} \dsub userm'(u).intPartner = \set{leave} \dsub userm(u).intPartner
%\end{fn}
%
%\plannote{(NPY) The version I want to keep is below.}
\begin{fn}{noModsToOthers}{userm, userm', good, sess} \\ 
\signature{(\mapof{Uid}{User}) \x (\mapof{Uid}{User}) \x \setof{Uid} \x Sid\to \Bool}
(\forall {u \in (\dom{userm} \minus good)}{ userm'(u) = userm(u)} ) \And \\
\forall*{u \in good}{
\set{sess} \dsub userm'(u).complete = \set{sess} \dsub userm(u).complete \And \\
\set{sess} \dsub userm'(u).intPartner = \set{sess} \dsub userm(u).intPartner}
\end{fn}

%\begin{lemma} \label{L-UnrealisticRely}
%Establishing that $post-NS1$ does not hold when using the previous specification of the sender and receiver with a more realistic rely condition. (I guess the one I put down as part of the NSL spec that follows below.)
%\end{lemma}

\caption{Specification for NS with users $from/to$ conforming to the protocol}
\label{F-spec-ideal}
\end{figure}

%%%%%%%%%%
\subsection{Pseudo-code}
\label{S-overall-spec}

The pseudo-code for NS is given in Fig.~\ref{F-dCode}, which uses statements with the following semantics:\\
$\kw{invent}: history :\sconc mk-Invent(me, n)$ ($n$ is completely new)\\
$\kw{send} (to, b) :\sconc mk-Msg(to, this, b)$\\
$\kw{rcv} \cdots$: picks up the most recent unread $Msg$ whose $rec$ field matches $this$. \\
Analogous with $x :+ 1$,
$s :\union t$ sets the final value of $s$ to $s \union t$, etc.

\begin{figure}
\begin{formula}
\begin{array}{l l}
	sender(to):
	&	receiver()\\
	\kw{wr } loc: User, s1:Sid 
	& 	\kw{wr } loc: User, s2: Sid\\
	s1 \gets \cdots
	&	s2 \gets \cdots \\
	loc.intPartner(s1) \gets to; \\
	NA \gets \kw{invent}; \\
	loc.knows(s1) :\union \set{NA};\\
	\kw{send} (to, \seq{this, NA}); \\
	&	\kw{rcv}(mk-Msg(this, , \seq{from, Nf})); \\
	&	loc.intPartner(s2) \gets from; \\
	&	loc.knows(s2):\union \set{Nf};\\
	&		NB \gets \kw{invent};\\
	&	loc.knows(s2):\union \set{NB};\\
	&		\kw{send} (from, \seq{Nf, NB}); \\
	\kw{rcv}(mk-Msg(this, , \seq{ret, Nt})) ; \\
	\kw{if } ret \neq NA \kw{ then abort };\\
	loc.knows(s1):\union \set{Nt};\\
	\kw{send} (to, \seq{Nt}); \\
	loc.complete(s1) \gets \true\\
	&		\kw{rcv}(mk-Msg(this, , \seq{ret})); \\
	&	\kw{if } ret \neq NB \kw{ then abort };\\
	&	loc.complete(s2) \gets \true
\end{array}
\end{formula}

%\plannote{(C) filling in the gap (see email of 25th) made me realise that $sess$ is not reflected here!?}

\caption{Distributed code for N-S}
\label{F-dCode}
\end{figure}
%Before getting to the distributed code in Fig.~\ref{F-dCode},
%it is useful to study the simple series of sequential steps in Fig.~\ref{F-sCode}
%which (given the rely condition%
%\footnote{Satisfied by the guarantee conditions of other instances of $NS$.})
%obviously achieve the post and guarantee conditions of $NS$
%(see Fig.~\ref{F-NS-spec}).
%Analogous with $x :+ 1$,
%$s :\union t$ sets the final value of $s$ to $s \union t$.
%
%\begin{figure}
%\begin{formula}
%NS1(from, to):\\
%\kw{ext wr } users: \mapof{Uid}{User}\T1
%	users(from).intPartner \gets to; \T1
%	\kw{let } NA \kw{ be s.t. }		NA \notin \Union \set{users(u).knows | u \in \dom{users}}\T1
%	users(from).knows :\union \set{NA};\T1  
%	users(to).intPartner \gets from; \T1
%	users(to).knows :\union\set{NA};\T1
%	\kw{let } NB \kw{ be s.t. }		NB \notin \Union \set{users(u).knows | u \in \dom{users}}\T1
%	users(to).knows :\union\set{NB};\T1
%	users(from).knows :\union\set{NB};\T1
%	users(from).complete \gets \true \T1
%	users(to).complete \gets \true
%\end{formula}
%\caption{Sequential pseudo-code for N-S}
%\label{F-sCode}
%\end{figure}

%\newpage
%%%%%%%%%%%%%%%%%%%%%%%
\subsection{Correctness argument}
\label{S-split1}

The correctness of the decomposition of $NS1$:

\begin{formula}
NS1 = sender(to) \parallel receiver() \parallel other
\end{formula}

\noindent
is interesting because there are actually two arguments:
synchronisation and that $sender$/$receiver$ are effectively running as disjoint parallel processes,
but the fact that the visibility of their nonces can be shown to be kept limited needs the assumptions~$no-leaks/no-forge$.
Although the assumptions match for $sender/receiver$,
these assumptions for $other$ are dangerous and are precisely the vulnerability of the N-S protocol.
Crucially, there can be many other messages than the intended three,
so we have to accept that these are a sub-sequence of the $history$ extension.

\begin{fn}{subseq}{s1, s2}
\signature{\seqof{X} \x \seqof{X} \to \Bool}
\exists{sel \in \seqof{\Bool}}{
\len{sel} = \len{s2} \And select(sel, s2) = s1}
\end{fn}

\begin{fn}{select}{sel, s}\\
\signature{\seqof{\Bool} \x \seqof{X} \to \seqof{X}}
\pre{\len{sel} = \len{s}}
\If sel = \emptyseq
\Then \emptyseq
\Else {
 \SIf \hd{sel} 
 \Then \seq{\hd{s}} \sconc select(\tl{sel}, \tl{s})
 \Else select(\tl{sel}, \tl{s})
 \Fi}
\Fi
\end{fn}

\begin{lemma} \label{L-synch}
Synchronisation of the two threads in Fig.~\ref{F-dCode} follows from the semantics of the message $\kw{send/rcv}$ statements. \end{lemma}
Proof. 
Providing there are no other messages,
the combined code would have the same effect as the obvious combined code.
If there is another message before the first $\kw{rcv}$ in $receiver$, it would initiate a separate session.
Because of the assumptions in $no-read-others$,
no $other$ user can generate a $Msg$ that is accepted by the $\kw{rcv}$ in $sender$.
The argument is similar for the second $\kw{rcv}$ in $receiver$.

\begin{lemma} \label{L-inv-pres}
The invariant $unique-nonces$ is preserved by both the $sender$ and $receiver$. \end{lemma}
Proof.
This holds because the chosen nonces are created by $\kw{invent}$.

The dynamic invariant $dyn-inv$ requires that $history$ is only extended, which is true of both processes.

\begin{lemma} \label{L-respect-guar}
The obligations in $inv-\Sigma$ regarding the (non) leakage of nonces is preserved by both the $sender$ and $receiver$. \end{lemma}
Proof. 
The $sender$ operation transmits a nonce ($Nt$) that it received in a $Msg$, but it returns it to the $Uid$ from where it appeared to have been sent, and similarly for $Nf$ in $receiver$.

\begin{lemma} \label{L-posts-achievable}
The obligations in $inv-\Sigma$ also include the (absence of) forgery of $Uid$s in $Msg$s. \end{lemma}
Proof. 
%\issue{Interestingly, we do need $no-forge$, for a while I doubted its need}
Only $sender$
(in pure $NS$)
includes a $Uid$ in the body of a $Msg$, which is not a forgery because it sends its own $Uid$ ($this$).

%
%\begin{lemma} \label{L-posts-achievable}
%The ($sender/receiver$) post conditions are satisfiable:
%because of the requirement that their nonces are not known by other than the intended users;
%this is where $no-leaks$ is essential.
%
%\issue{redo}
%\end{lemma}

%\begin{lemma} \label{L-post-NS}
%Establishing that $post-NS$ holds
%(writing $post-sender/receiver$ for the (obvious?) combination of the two sub-operations).
%
%\begin{formula}
%post-sender(\sigma, \sigma') \And post-receiver(\sigma, \sigma') \Implies post-NS(\sigma, \sigma')
%\end{formula}
%\end{lemma}

%\discuss{OK there's detailed checking to be done here, but I think it's close to provable}

%\plannote{(C) We need the key Lemma on the secrecy of $NA, NB$! \\
%(NPY) I'm confused - isn't that what's now in the post condition?}

\begin{lemma} \label{L-nonce -secrecy}
$post-NS$ requires that only $from/to$ can have $NA/NB$ in their $knows$ fields.

Proof:
Since $from/to$ both have $conforms$, neither can leak the nonces;
it follows from $no-read-others$ that there is no other way for a mischievous user $u$ to access the nonces 
and $unique-nonces$ ensures that identical nonces cannot be generated.
\end{lemma}

%%%%%%%%%%%
\subsection*{Fault tolerance}

\begin{lemma} \label{L-fail-NS}
The flag $complete$ is not set to $\true$ if any errant messages interfere
(i.e. $sender$ or $receiver$ abort).
\end{lemma}

\section{Lowe's analysis}
\label{S-Lowe}

Lowe's counterexample, described in \S\ref{S-Lowe-attack}, shows the need for precise specifications;
Lowe's fix is easy to represent in the ``distributed code'' with only a few minor changes
to the original N-S version in Fig.~\ref{F-dCode}.
%:one extra field in message $b1$ and an extra if/then/else in $sender$.
%(for consistency, should we also add $A$ to message $a2$?) \\
Lowe's fix is also correct with respect to $post-NS$. However, the assumptions embodied in $inv-\Sigma$
are unrealistic and cannot be satisfied in the presence of miscreants; \S\ref{S-Lowe-check} addresses this
by offering a different post condition, $post-NS(L)-FT$, that holds under more relaxed and realistic assumptions. 
%(in other words, he hasn't messed anything up)\\
%- $inv-\Sigma$  (the effective assumption of $NS$) fails in Lowe's counter example: message $b1'$ triggers an abort\\
%- but that's not a proof that there are no other attacks!\\
%- the real issue is to find a specification which is not just a description of Lowe's counter example
%(this is the thrust of \S\ref{S-Lowe-assms})
%but it is different from $post-NS$
More importantly,
the assumptions are spelled out.

%%%%%%%%%%%%%%%%%%%%%%%%%
\subsection{Lowe's attack and ``correction''}
\label{S-Lowe-attack}

%\issue{Again I'm experimenting with moving this}

The N-S protocol was accepted as being correct for 18 years before Lowe discovered an attack \cite{lowe1995attack}.
At first sight, Lowe's attack: 

\begin{formula}
(a1) A: enc([A, NA], pkeym(I))\\
(d1) I: enc([A, NA], pkeym(B))\\
(b1) B: enc([NA, NB], pkeym(A))\\
%(d2) I: \kw{skip}\\
(a2) A: enc([NB], pkeym(I))\\
(d3) I: enc([NB], pkeym(B))
\end{formula}

\noindent
is odd in that $A$ starts by communicating with a partner who does not conform to the protocol, 
referred to as $I$ (for intruder) in the following. It is important to accept that $A$ intends
to communicate with $I$, unlike man-in-the-middle attacks, though $A$ does not know that $I$ is dishonest.%
%(but if there were a predicate to determine if a user was honest, life would be simple).
\footnote{It is paradoxical that it is $a2$ that completes the intruder's knowledge
since this message was clearly intended as a confirmation.}
It is clear that $I$ immediately violates both of the earlier rely conditions  
($no-leaks, no-forge$).
Unfortunately, it is not possible to check that a recipient does not ``leak'' secrets
(and it would also be possible for $sender$ to leak the nonces).

Notice that Lowe's attack is not a standard ``man in the middle'' anomaly
in which the duped user has a separate session with the intruder.
In Lowe's attack,
$B$ has evidence that their partner is $A$;
the danger here is that $B$ accepts this fact to release some information and/or take some action on the basis that $A$ is a trusted user.

The defence that Lowe suggests results in a more elaborate protocol 
(to which both partners have to conform) where all messages include the (encrypted) identity of the sender;
this makes it possible to detect Lowe's intruder: 

\begin{formula}
(a1) A: enc([A, NA], pkeym(I))\\
(d1) I: enc([A, NA], pkeym(B))\\
(b1') B: enc([B, NA, NB], pkeym(A))\\
(a-abort) \\
A \hbox{ detects the problem because }[I, NA, NB]  \hbox{ was expected.} 
\end{formula}

Notice that these changes are straightforward to reflect in the code shown in Fig.~\ref{F-dCode}.

At first glance, Lowe's defence appears to rely on a relatively dangerous assumption, that if a user sends their identity, it really is their $Uid$
(i.e. $no-forge$);
%
%\begin{fn}{rely-L}{mk-Msg(s, vl, rec)}
%\signature{Msg \to \Bool}
% u \in (Uid \inter \elems{vl}) \Implies  u = s
%\end{fn}
%
%\issue{= $no-forge$?}
%
this is actually broken at $d1$ by $I$, but its after-effects are detected at $b1'$.
However, the clever aspect of Lowe's correction is in fact that it works \textit{despite} the fact that dishonest users could be lying about
their identity.

%%%%%%%%%%%%%%%%%%%%%%%%%
\subsection{Specifying the extra check}
\label{S-Lowe-check}

The idea hinted at in \S\ref{S-RG} of specifying software by layers of R-G conditions can be applied to security protocols.
In the case of Lowe's extension to N-S
(henceforth referred to as NSL):

\begin{itemize}

%\plannote{(C) We  should say here that Lowe's extension still satisfies Fig. \ref{F-spec-ideal}.}
\item The extended protocol above satisfies $post-NS$, as in Fig.~\ref{F-spec-ideal}, under the optimistic assumptions recorded in $inv-\Sigma$;

\item Lowe's extension does not attempt to deliver agreement between $from$ and $to$ with weaker --more realistic-- assumptions

\item What the extra self-identity in message $b1$ facilitates is an extra test by $sender$, which can detect that the message $a1$ is actually coming from a different user, rather than the user to
whom the message was sent ---
this indicates that there is a miscreant involved.

\end{itemize}

Therefore, the second layer of the R-G specification must indicate that the protocol should terminate abnormally in this case. This leads to the following specification:

\begin{op}[NS(L)-FT]
\args{from: Uid, to: Uid, sf: Sid, st: Sid}
\ext{\Wr users: \mapof{Uid}{User} \\
        \Wr history: \seqof{Action}}
\pre{users(from).conforms \And \\
\Not users(from).complete(sf) \And \Not users(to).complete(st)}
\rely{noMods(users, users', from, sf) \And noMods(users, users', to, st)}
\guar{noModsToOthers(users, users', \set{from, to})}
\post{
 \exists*{NA \in users'(from).knows(sf)}{
        \exists*{u \in (Uid \minus \set{from, to}), su \in Sid}{
	users'(u).conforms \And\\
	NA \in users'(u).knows(su) \And\\
	users'(u).intPartner(su) = from \Implies\T5
		\Not users'(from).complete(sf)
		}
      }
      }
\end{op}
Assuming that $from$ is conforming, this post condition ensures that there cannot be a third user who is
conforming, has knowledge of $NA$ and has an $intPartner$ of $from$. 
It is important to recognise that a simple ``man in the middle'' attack can still work
(it would satisfy a specification which omitted $users'(u).intPartner(su) = from$).

%\plannote{(C) Here's the biggest change (work to be done):\\
%As per my email of 26th,
%I suspect that both FT1/2 are wrong to say in their post conditions that $from/to$ have $complete = \true$!\\
%To me, Lowe doesn't aim to complete successfully in a less honest environment ---
%instead, it promises to abort with certain detectably dishonesty.\\
%So  I think the overall form of $post-NSL-FT$ should include:
%\begin{formula}
%\exists*{u \in (Uid \minus \set{from, to}), su \in Sid}{
%	users'(u).conforms \And\\
%	\set{NA, NB} \subseteq users'(u).knows(su) \Implies\T5
%		\Not users'(from).complete(sf)
%		}
%\end{formula} 
%(NPY) This isn't correct. There can be another user who has NA and NB. Recall my counterexample: \\
%A sends (A,NA) to I. \\
%I sends (I,NA) to B. \\
%B sends (B,NA,NB) to I. \\
%I sends (I,NA,NB) to A. \\
%All three are happy and all three have NA and NB. \\
%(C) Of course, my reservation about Lowe's modification is that it feels like one specific identified dishonesty}

%\plannote{(C) I also think that we'll have to remove $no-leaks/no-forge$ from $inv-\Sigma$ and have them in R-Gs}

%\plannote{(C) I'm even wondering whether we shouldn't just specify the behaviour of $from$? \\
%Chat about this because I don't understand.}

Notice that an implementation of a layered R-G specification must 
(be shown to)
satisfy all layers, so when both users are conforming, the requirements of $post-NS(L)$ in Fig. \ref{F-spec-ideal} must be satisfied as well.

\section{Encryption introduced as a reification}
\label{S-reif}

%\plan{I think this section could be reduced to a single paragraph?}
The notion that only the intended user 
($rec$ field)
can access the content of a $Msg$ can be achieved using encryption.
(This idea is also used in~\cite{sprenger2010developing}.)
In our work 
%(omitted here for space reasons) 
we have been careful to identify the extra assumptions that are needed.

Sections~\ref{S-postpone-encryption} and~\ref{S-Lowe} identify crucial assumptions about the content of messages;
this section outlines how rely-guarantee conditions can be used to argue that encryption can discharge the secrecy assumptions on the $rec$ field of $Msg$;
this brings in additional assumptions about $SKey/PKey$.

%\begin{record}{\Sigma\sb{L}}
% msgs: \seqof{EncAction}\\ 
% \end{record}
% \where
% \begin{fn}{inv-\Sigma\sb{L}}{mk-\Sigma\sb{L}(as)}
% no-spy(as)
% \end{fn}

%\plannote{(C) Oops: we've lost the definition of $Local$;
%it needs the $SKey$ and invariants thereon;
%I'll hunt after reading \S\ref{S-states}}

%%%%%%%%%%%%
%\subsubsection{Public/secret keys} 

%\type{EncAction}{EncMsg | Invent}

%\begin{record}{EncMsg}
%      send: Uid  \hspace{10em} \hbox{ghost variable} \\
%      content: \seqof{Item} \\
%      encFor: Uid
%\end{record}

%\begin{record}{PlainMsg}
%      send: Uid  \hspace{10em} \hbox{ghost variable} \\
%      content: \seqof{Item}
%\end{record}

Messages are encrypted lists of items that are either $Nonce$ text or user identifiers.
The essence of public key encryption is the use of matching keys:

\begin{formula}
match: PKey \x SKey \to \Bool
\end{formula}

\noindent
Assumptions on $match$ need to be added to $inv-\Sigma$.
%\issue{additional rely conditions:\\
%can't read another $User$ (especially $skey$)
%}
Furthermore,
frame descriptions must mark that no user can read the secret key of another user.

Encryption/decryption functions are used: 

\begin{formula}
enc: \seqof{Item} \x PKey \to EncMsg\\
dec: EncMsg \x SKey \to \seqof{Item}
\end{formula}

\noindent
These are related by the crucial $dec-enc-prop$  property:

%\plannote{(N) named property; (C) after more thought, I've left it as a global property:
%--- strictly it doesn't only apply to $Msg$ in $\Sigma$}

\begin{formula}
\forall{cm \in \seqof{Item}}{
  dec(enc(cm, pk), sk) = cm \Iff match(pk, sk)}
\end{formula}

\noindent
Decrypting an encrypted message using the correct secret key (and only that key) should return valid information:
\begin{formula}
  \forall{m \in EncMsg}{
  dec(m, sk) \in \seqof{Item} \Iff \exists{pk \in PKey}{match(pk, sk)} }
\end{formula}

\begin{formula}
  \forall*{sk \in SKey, pk \in PKey}{match(pk, sk)  \Iff \exists{u \in Uid}{u.skey = sk \And pkeys(u) = pk}} 
\end{formula}

%%%%%%%%%%%%%%%%
\section{Conclusion and Related Work}

%\plannote{Putting anything here shifts the section title back to the previous page!\\
%It's ok now since it's now made the previous section have better spacing!}

We have sought to illustrate how a state-based notation can expose interactions between processes.
We feel that,
in contrast to process algebraic notations,
focussing on the interference can highlight issues that in turn can be recorded and reasoned about using rely-guarantee conditions.

The importance of clearly identifying assumptions in models of security protocols has been identified by several researchers. Halpern and Pucella note that the Dolev-Yao model is insufficient in many cases \cite{HalpernPucella12};
to identify certain kinds of attacks, different adversary models are required, which place different
assumptions on the environment. 
%They point out that even the verifiers which allow the most expressivity
%in describing the attacker capabilities are still restricted to specific adversary models. 
Halpern and
Pucella also state: 
\textit{The problem is even worse when it is not clear
exactly what assumptions are implicitly being made about the adversary.} 
Armando et al. \cite{ArmandoCC09} support the view that realistic assumptions need to be made in order to
ensure that a protocol is ready for deployment, commenting: 
\textit{Most model checking techniques for security protocols make a number of simplifying
assumptions on the protocol and/or on its execution environment that greatly complicate or even
prevent their applicability in some important cases.} 
This view is shared by Bond and Clulow 
\cite{BondC05}, who express concern that protocols verified by formal tools are restricted by unrealistic
assumptions, and protocol designs based on these assumptions are \textit{over-engineered and impractical to deploy}. Pass \cite{Pass11} lists several well-known protocols which have been proven correct under certain
assumptions, and questions whether these protocols would still be secure under a different set of assumptions. 
%For example, Pass investigates whether the Schnorr identification scheme, which was shown to be
%correct under passive attacks, would still hold under active attacks.

%Research on security protocol verification has followed two main streams: symbolic approaches, in
%which a formal model of the protocol is used which hides details such as the cryptographic algorithms, and
% computational approaches, which reason about the mathematical properties of the cryptographic algorithms
% and include a notion of probability. While computational approaches are more realistic, symbolic methods have the advantage of being
% less complex, allowing a verification result to be produced in much less time.
% Abadi and Rogaway \cite{AbadiRogaway07} give the first attempt at relating
% the two approaches, demonstrating that it is possible to find a relationship between them. Coutier et al. \cite{CortierKW11} survey the attempts to relate the two approaches. 
Formal approaches for protocol verification follow two main streams: symbolic and computational approaches. Symbolic approaches typically utilise model checking, using tools such as ProVerif \cite{Blanchet13}, Scyther \cite{Cremers08} and OFMC \cite{BasinMV05}. CryptoVerif \cite{Blanchet23} is an example of a tool based on a computational approach but with symbolic aspects. For a comprehensive survey of formal approaches for protocol verification see \cite{kulik2022survey}. However, there has been very little effort in
finding methods for clearly identifying and articulating assumptions, which is the main advantage of the rely/guarantee approach presented in this paper.

Future work includes the use of tool support to mechanise the descriptions, 
but Overture
(see https://www.overturetool.org/)
could require extensions to handle rely-guarantee conditions.

%%%%%%%%%%%%%%%%%%%%%%%%
\nocite{BowenEt-23}

\bibliography{parallel}

\end{document}